\documentclass[a4paper,11pt]{article}
\pdfoutput = 1
\usepackage{jheppub}
\usepackage{amsmath,amssymb,amscd,braket,amsfonts}
\usepackage{color}
\usepackage{graphicx}
\usepackage{slashed}
\usepackage{amsthm}

\usepackage{amstext} % for \text macro
\usepackage{array}   % for \newcolumntype macro
\newcolumntype{L}{>{$}l<{$}} % math-mode version of "l" column type

\def\p{\partial}
\newcommand{\bea}{\begin{eqnarray}}
\newcommand{\eea}{\end{eqnarray}}
\newcommand{\be}{\begin{equation}}
\newcommand{\ee}{\end{equation}}
\newcommand{\ba}{\begin{align}}
\newcommand{\ea}{\end{align}}

\newcommand{\re}{\mathfrak{Re\,}}

\title{QNM orthogonality relations for AdS black holes}
\date{2025}

\author{Paolo Arnaudo,}
\author{Javier Carballo,}
\author{and Benjamin Withers}
\affiliation{Mathematical Sciences and STAG Research Centre, University of Southampton, Highfield, Southampton SO17 1BJ, UK}
\emailAdd{p.arnaudo@soton.ac.uk}
\emailAdd{j.carballo@soton.ac.uk}
\emailAdd{b.s.withers@soton.ac.uk}

\abstract{We present orthogonality relations for quasinormal modes of a wide class of asymptotically AdS black holes. The definition is obtained from a standard product, modified by a CPT operator and placed on a complex radial contour which avoids branch points of the modes. They are inspired by existing constructions for de Sitter and Kerr spacetimes. The CPT operator is needed to map right eigenfunctions of the Hamiltonian into left eigenfunctions. The radial contour connects two copies of the dual QFT on a thermal Schwinger-Keldysh contour, making contact with real-time holography and the double cone wormhole.}

\begin{document}
\maketitle

\section{Introduction}
Quasinormal modes (QNMs) provide a universal characterisation of the decay of black holes towards equilibrium at asymptotically late times.  This makes them an invaluable tool in the theoretical study and gravitational wave observations of dynamical black holes. Their decay is due to a dissipative linear-response process in which energy falls through $\mathcal{H}^+$ or is radiated to $\mathcal{I}^+$. However, this dissipative process means that QNMs are not orthogonal to each other in any standard way.

Despite this, orthogonality relations between de Sitter static patch QNMs have been constructed by applying suitable discrete symmetry operations to the Klein-Gordon product \cite{Jafferis:2013qia}, building on \cite{Witten:2001kn, Bousso:2001mw, Ng:2012xp}. Here the radial integral in the product avoids QNM singularities by appropriate $i\epsilon$ insertions. See also \cite{Crawley:2021ivb, Cotler:2023qwh} for analogous constructions in celestial holography.

More recently, a similar construction was presented for Kerr spacetimes, employing $\mathcal{T}\varphi$-reflections \cite{Green:2022htq} with applications in \cite{Cannizzaro:2023jle}. In this construction the approach begins with a radial integral which is divergent, and then regulated \cite{PhysRevA.49.3057, Ching:1995rt}. Relatedly, in \cite{London:2023aeo, London:2023idh} appropriate integral weights for orthogonality products were identified.  In \cite{Zhu:2023mzv} orthogonality relations for discretised systems are also constructed.

\begin{figure}[h!]
\centering
\includegraphics[width=\columnwidth]{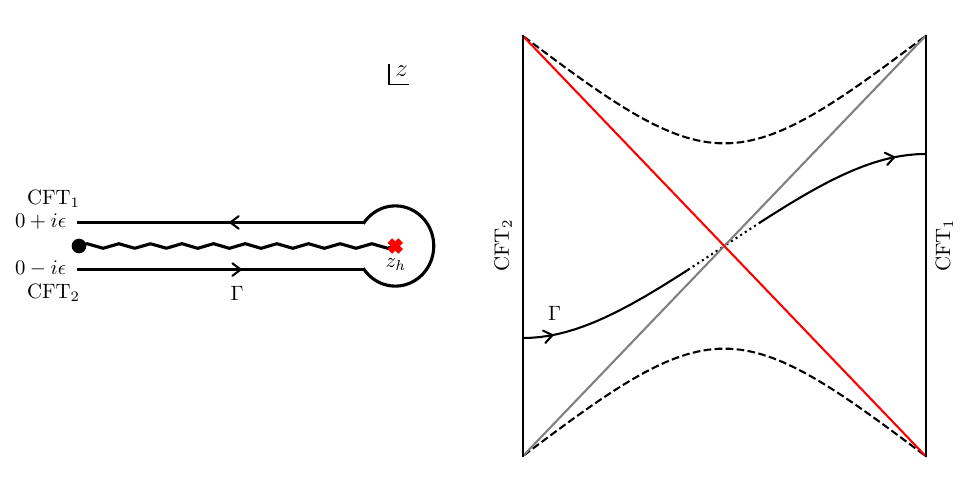}
\caption{\textbf{Left:} The complex radial contour $\Gamma$ used in our QNM orthogonality product. The contour avoids branch point singularities of QNMs and anti-QNMs on the horizon. \textbf{Right:} The contour $\Gamma$ on the maximally extended spacetime, connecting two copies of the QFT through Euclidean time translation by half the thermal circle, $\beta/2$. Here the red line illustrates a singular past horizon on side 1, corresponding to the branch point singularities of QNMs on that side, avoided by the $i\epsilon$ insertions.}
\label{fig:contour}
\end{figure}

In this work, inspired by the above examples, we construct new orthogonality relations for QNMs of a wide class of asymptotically AdS black holes. Our orthogonality relation similarly employs discrete symmetry transformations to modify an existing inner product, and are defined on a complex radial contour. The contour encircles the QNM branch points on the horizon and connects two AdS boundaries, see figure \ref{fig:contour}. Such contours naturally arise in the context of real-time AdS/CFT \cite{Herzog:2002pc, Skenderis:2008dh, Skenderis:2008dg, Son:2009vu, deBoer:2018qqm, Glorioso:2018mmw, Jana:2020vyx}. In the bulk they can be interpreted as a slice of the maximally extended black hole spacetime, with appropriate $i\epsilon$ insertions that avoid QNM singularities on the horizon. In the holographic dual they connect two copies of the boundary QFT on a thermal Schwinger-Keldysh contour. This contour also appears in the context of the double cone wormhole \cite{Saad:2018bqo}.

Let us first illustrate our results in the simple case of a Klein-Gordon scalar $\Phi$ of mass $m_\Phi^2 = \Delta(\Delta-2)$ on the non-rotating BTZ black hole \cite{Banados:1992wn},
\be
ds^2=\frac{1}{z^2}\left(-(1-z^2)dt^2+\frac{1}{(1-z^2)}dz^2+d\varphi^2 \right), \label{BTZmetric}
\ee
with AdS boundary at $z=0$ and event horizon at $z=1$.
The Klein-Gordon product defined on the contour $\Gamma$ is given by
\be
\braket{a,b}_\text{KG} = i \int_{0}^{2\pi} d\varphi\,\int_\Gamma dz\,\frac{1}{z(1-z^2)}\,a^*\overleftrightarrow{\partial_t} b, \label{KGnormBTZ} 
\ee
where $a^*\overleftrightarrow{\partial_t} b \equiv a^*\p_tb-(\p_ta^*)b$. This is the integral of the discontinuity of the integrand due to branch points in the fields. Given the Hamiltonian $\mathcal{H} = i \partial_t$, under  \eqref{KGnormBTZ} one has the following standard orthogonality relation between its left and right eigenfunctions, $u_i$ and $v_j$ respectively,
\be
\braket{u_i, v_j}_\text{KG} \propto \delta_{ij}, \label{leftright}
\ee
where $\mathcal{H}v_i = \omega_i v_i$, and $\mathcal{H}^{\dagger_\text{KG}}u_i = \omega^*_i u_i$, where $\mathcal{H}^{\dagger_\text{KG}}$ denotes the adjoint under \eqref{KGnormBTZ}, i.e. $\braket{\mathcal{H}^{\dagger_\text{KG}}a,b}_\text{KG} = \braket{a,\mathcal{H}b}_\text{KG}$. This follows from
\be
\omega_j \braket{u_i, v_j}_\text{KG} = \braket{u_i, \mathcal{H}v_j}_\text{KG} = \braket{\mathcal{H}^{\dagger_\text{KG}} u_i, v_j}_\text{KG} = \omega_i\braket{u_i, v_j}_\text{KG},
\ee
so that if $\omega_i \neq \omega_j$ then $\braket{u_i, v_j}_\text{KG} = 0$.
Our orthogonality relation is inherited from the standard relation \eqref{leftright} by providing a mapping between left and right eigenfunctions. The key point is that $\mathcal{H}^{\dagger_\text{KG}} = \mathcal{H}$, which can be shown as follows. By using the equation of motion and integration by parts on the contour $\Gamma$ one obtains
\be
\braket{a,\mathcal{H}b}_\text{KG} = \braket{\mathcal{H}a,b}_\text{KG} + \int d\varphi  \left[\frac{1-z^2}{z}\left(b\partial_za^* - a^*\partial_z b\right)\right]_{z=0-i \epsilon}^{z=0+i\epsilon},\label{KGHermitian}
\ee
thus for normalisable perturbations, that is, $a,b \propto z^{\Delta}$ as $z\to 0$ with $\Delta > 1$, the boundary terms vanish. Since $\mathcal{H}^{\dagger_\text{KG}} = \mathcal{H}$, any right eigenfunction is also a left eigenfunction, so we can put right eigenfunctions in both slots, $\langle v_i,v_j\rangle_{\text{KG}}$, and still take advantage of \eqref{leftright}. However, unlike \eqref{leftright}, the matrix $\langle v_i,v_j\rangle_{\text{KG}}$ is not diagonal.\footnote{The vanishing of the Klein-Gordon norms (i.e. the diagonal entries of $\langle v_i,v_j\rangle_{\text{KG}}$) is precisely what permits complex eigenvalues even though $\mathcal{H}^{\dagger_\text{KG}} = \mathcal{H}$. The problem with using the Klein-Gordon norm for QNMs was pointed out before \cite{Jafferis:2013qia}.} 
Diagonalising the Klein-Gordon product leads us to our main definition, used throughout this work,
\be
\braket{a,b} \equiv \braket{\mathcal{CPT}a,b}_\text{KG},\label{orthnorm}
\ee
where $\mathcal{C}$ denotes complex conjugation, $\mathcal{T}$ denotes $t\to -t$, and in this particular case of BTZ, $\mathcal{P}$ denotes $\varphi\to -\varphi$. The role of $\mathcal{CPT}$ is to map an eigenfunction of $\mathcal{H}$ with eigenvalue $\omega$ into an eigenfunction with eigenvalue $\omega^*$, and hence $\braket{v_i,v_j} = \braket{u_i, v_j}_\text{KG} \propto \delta_{ij}$, as desired. This property of $\mathcal{CPT}$ can be seen from the separated form of eigenfunctions and commuting $\mathcal{CPT}$ with the equation of motion. Finally, in section \ref{sec:efuncs}, we show that the regular, normalisable eigenfunctions of $\mathcal{H}$ on $\Gamma$ are the QNMs and anti-QNMs, completing the proof that \eqref{orthnorm} provides the desired orthogonality relations among these modes.

We can also compute \eqref{orthnorm} on pairs of eigenfunctions to explicitly demonstrate their orthogonality. 
The BTZ QNMs $\Phi_{nm}^\pm(t, z, \varphi)$ are given in \eqref{QNMBTZ}, while the anti-QNMs $\widetilde{\Phi}_{nm}^\pm(t, z, \varphi)$ are given in \eqref{aQNMBTZ}, related by $\mathcal{CPT} \Phi_{nm}^\pm = \widetilde{\Phi}_{nm}^\pm$. They are labeled by an azimuthal quantum number $m$, a parity sector $\pm$, and an overtone number $n$. The computation presented in appendix \ref{sec:BTZ} shows, 
\bea
\braket{\Phi_{nm}^\pm,\Phi_{n'm'}^{\pm'}} &=& \frac{(2 \pi)^2 i   (-1)^{n-1}\, n! \,\Gamma \left(\Delta\right)}{\Gamma \left(\mp i m-n\right)\, \Gamma \left(n\pm i m+\Delta\right)\,\left(\Delta\right)_{n}\,\left(\Delta \pm i m+2n\right)}\,\omega^{\pm}_{nm}\,\delta_{mm'}\,\delta_{nn'}\delta_{\pm\pm'},\qquad\; \label{BTZoverlap1}\\
\braket{\widetilde{\Phi}_{nm}^\pm,\widetilde{\Phi}_{n'm'}^{\pm'}} &=& \frac{(2 \pi)^2 i   (-1)^{n-1}\, n! \,\Gamma \left(\Delta\right)}{\Gamma \left(\pm i m-n\right)\, \Gamma \left(n\mp i m+\Delta\right)\,\left(\Delta\right)_{n}\,\left(\Delta \mp i m+2n\right)}\,\widetilde{\omega}^{\pm}_{nm}\,\delta_{mm'}\,\delta_{nn'}\delta_{\pm\pm'},\label{BTZoverlap2}\\
\braket{\Phi_{nm}^\pm,\widetilde{\Phi}_{n'm'}^{\pm'}} &=& 0,\label{BTZoverlap3}
\eea
where $\omega^{\pm}_{nm}$ and $\widetilde{\omega}^{\pm}_{nm}$ are the associated eigenfrequencies \eqref{wBTZ}, \eqref{awBTZ}. The result demonstrates orthogonality between all modes, in accordance with the above proof.\footnote{Note here the important caveat that the $m=0$ modes have zero norm. This can be understood a degeneration between the orthogonal $\pm$ sectors at $m=0$.}
\\\\
The layout of the rest of the paper is as follows. In section \ref{sec:orthogonality} we extend the BTZ example presented above to a general class of AdS$_{d+1}$ black hole spacetimes, with general metric functions required to obey only certain properties. In particular we prove orthogonality of the regular, normalisable eigenfunctions of $\mathcal{H}$ on $\Gamma$ for this class of spacetimes. In section \ref{sec:efuncs} we prove that the regular, normalisable eigenfunctions of $\mathcal{H}$ on $\Gamma$ are, indeed, just the QNMs and anti-QNMs of these general black holes. In section \ref{sec:examples} we turn to explicit numerical evaluation of \eqref{orthnorm} with QNMs and anti-QNMs in the slots for Schwarzschild-AdS$_4$, confirming the result. We finish with a discussion in section \ref{sec:discussion}.

\section{Orthogonality for a general class of black holes}\label{sec:orthogonality}
In this section we generalise the BTZ results from the introduction to a general class of asymptotically-AdS$_{d+1}$ black hole spacetimes, given by the following line element,
\be
ds^2 = \frac{1}{z^2}\left(-f(z)dt^2+\frac{dz^2}{g(z)}+d\sigma_{d-1}^2 \right), \label{generalBG}
\ee
where $d\sigma_{d-1}^2$ is the spatial geometry with constant sectional curvature $K = 0,\pm1$. The conformal boundary is reached as $z\to 0$ near which the metric functions behave as
\be
f(z) = 1 + O(z), \qquad g(z) = 1 + O(z),
\ee
and there is an outer, non-extremal event horizon at $z=z_h$ so that
\be
f(z) = f_0(z-z_h) + O(z-z_h)^2, \qquad g(z) = g_0(z-z_h) + O(z-z_h)^2.
\ee
Such black holes have a temperature,
\be
\beta^{-1} = \frac{\sqrt{f_0\,g_0}}{4\pi}.
\ee
This includes a wide class of symmetric black hole spacetimes including Schwarzschild-AdS$_{d+1}$ for which $f(z) = g(z) = 1+ Kz^2-(1+Kz_h^2)\frac{z^d}{z_h^d}$, as well as other black holes with matter, such as Reissner-Nordstrom-AdS$_{d+1}$.

We consider a complex scalar $\Phi$ perturbation on the background \eqref{generalBG} with mass given by $m_\Phi^2=\Delta(\Delta-d)$ and we restrict our attention to the larger root of this equation, so that $2\Delta > d$. 
The Klein-Gordon equation is
\be
i\partial_t \begin{pmatrix} 
   \Phi \\
     \partial_t \Phi 
    \end{pmatrix}=\mathcal{H}\,\begin{pmatrix} 
   \Phi \\
     \partial_t \Phi
    \end{pmatrix}\,,\quad \mathcal{H}=\begin{pmatrix}
0 & i\\
\mathcal{L}& 0
\end{pmatrix}
\label{genKG}
\ee
with 
\be
\mathcal{L} = ifg\partial_z^2 + i \frac{zgf'+f(-2(d-1)g + z g')}{2z}\partial_z + i f\Delta_{\sigma}-i\frac{m_\Phi^2 f}{z^2}, \label{Lop}
\ee
where $\Delta_{\sigma}$ is the spatial Laplacian. The Klein-Gordon product, defined on the contour $\Gamma$, is given by
\bea
\braket{a,b}_\text{KG} &\equiv& i \int d\Sigma^\mu\, a^*\overleftrightarrow{\partial_\mu} b\\
&=& i \int d\sigma_{d-1} \int_\Gamma dz\,\frac{1}{z^{d-1}\sqrt{f(z)}\sqrt{g(z)}}\,a^*\overleftrightarrow{\partial_t} b,
\eea
which we use to define our bilinear form $\braket{\cdot,\cdot}$ through the relation \eqref{orthnorm}, inserting $\mathcal{CPT}$ into the left slot. $\mathcal{CPT}$ is required for the reasons discussed in the introduction; without it, one does not obtain the correct pairings of left and right eigenfunctions, which leads to zero norms. As before in the BTZ case, $\mathcal{C}$ corresponds to complex conjugation, and $\mathcal{T}$ to time reversal. However, $\mathcal{P}$ depends on the choice of $K$, the basic requirement is that $\mathcal{CP}$ should leave eigenfunctions of $\Delta_\sigma$ invariant. For $K=0$, $\mathcal{P}$ corresponds to a parity operator, $\sigma \to -\sigma$, for all spatial coordinates $\sigma$. For $K=1$, $\mathcal{P}$ corresponds to $\varphi\to -\varphi$ where $\varphi$ is the azimuthal angle on the sphere.

Finally we come to the demonstration of orthogonality under \eqref{orthnorm}, which parallels that of the BTZ case in the introduction.
Using \eqref{genKG} and integration by parts on the contour $\Gamma$, one has,
\bea
\braket{a,(\mathcal{H} - \mathcal{H}^\dagger) b}  &=& \int d\sigma_{d-1}  \left[z^{1-d}\sqrt{f(z)}\sqrt{g(z)}\left(b\partial_z(\mathcal{PT}a) - (\mathcal{PT}a)\partial_z b\right)\right]_{z=0-i \epsilon}^{z=0+i\epsilon}.
\eea
Normalisable perturbations obey $a\sim a_\Delta z^\Delta$, $b\sim b_\Delta z^\Delta$ for $z\to 0$, and so the integrand behaves as $\sim z^{2\Delta -d}$ for $z\to 0$. Thus we recover that $\mathcal{H}^\dagger = \mathcal{H}$, with respect to the bilinear form \eqref{orthnorm}.\footnote{Just as in the introduction, $\mathcal{H}^\dagger = \mathcal{H}$ does not imply reality of eigenvalues, here it is because \eqref{orthnorm} is linear in both slots, rather than linear in one and anti-linear in the other.} Note that $\mathcal{H}^\dagger=\mathcal{H}$ also follows immediately from $\mathcal{H}^{\dagger_{\text{KG}}}=\mathcal{H}$ given that $[\mathcal{H},\mathcal{CPT}]=0$. It then follows that regular, normalisable eigenfunctions of $\mathcal{H}$ with different frequencies $\omega$ are orthogonal to one another under \eqref{orthnorm}, since if $v_i, v_j$ are such eigenfunctions with eigenvalues $\omega_i,\omega_j$, then
\be
(\omega_j - \omega_i)\braket{v_i,v_j} = \braket{v_i,\mathcal{H}v_j}-\braket{\mathcal{H}v_i,v_j} = 0,
\ee
so that if $\omega_i \neq \omega_j$ then $\braket{v_i,v_j}= 0$.

\section{Regular, normalisable eigenfunctions of $\mathcal{H}$ on $\Gamma$}\label{sec:efuncs}
In the previous section we showed that regular, normalisable eigenfunctions of $\mathcal{H}$ on $\Gamma$ are all orthogonal to each other under \eqref{orthnorm}. In this section we will show that such eigenfunctions correspond to the QNMs and anti-QNMs of the black hole. An analogous result can be found in the context of the double cone wormhole \cite{Saad:2018bqo, Bah:2022uyz, Chen:2023hra}.\footnote{We thank V. Ziogas for discussions on this point.}

We are looking for eigenfunctions of $\mathcal{H} = i \partial_t$. These take the form
\be\label{separation}
\Phi(t,z,\sigma) = e^{-i\omega t} S(z,\sigma)
\ee
where $\omega$ is the eigenvalue. Without loss of generality we can decompose $S$ as follows,
\be
S(z,\sigma) =  \sum_J Z_J(z)Y_J(\sigma)
\ee
where $Y$ are eigenfunctions of $\Delta_\sigma$ labeled by quantum numbers $J$. Then $Z_{J}(z)$ obeys
\be
\mathcal{L}_J Z_{J}(z) =-i\omega^2 Z_{J}(z), \label{Zeq}
\ee
where $\mathcal{L}_J$ is the operator \eqref{Lop} with $\Delta_\sigma$ replaced by its eigenvalue, $-\lambda_J$ where $\Delta_\sigma Y_J = -\lambda_J Y_J$. From here on we will drop the label on $Z_{J}$. 
Near the horizon, the general solution to \eqref{Zeq} takes the form\footnote{In cases where the powers are separated by an integer -- i.e. at (Wick rotated) Matsubara frequencies $\omega_n = \frac{2\pi n}{\beta}i$ -- there are also logarithms.}
\bea\label{nearhorizon}
Z(z) = \hat{c}_A(z-z_h)^{\frac{i\beta\omega}{4\pi}}(1 + \ldots) + \hat{c}_R(z-z_h)^{- \frac{i\beta\omega}{4\pi}}(1 + \ldots),
\eea
here the cuts are arranged to the left of $z_h$, towards the boundary.
 The prefactor $\hat{c}_A$ multiplies an outgoing field, and the solution integrated from the horizon to the boundary (on either sheet) is the advanced bulk-to-boundary Green's function, $\widetilde{G}_A(\omega,z)$. Similarly, $\hat{c}_R$ multiplies ingoing behaviour and in integrated form is the retarded bulk-to-boundary Green's function, $\widetilde{G}_R(\omega,z)$, hence,
\be
Z(z) = c_A \widetilde{G}_{A}(\omega,z) + c_R\widetilde{G}_{R}(\omega,z),\label{ZGreens}
\ee
where the coefficients have been adjusted to incorporate our normalisation choice, $\widetilde{G}_{A,R}(\omega,z) \sim z^{d-\Delta}$ as $z\to 0$.
In our setup, we have two copies of the (real) spacetime, which we can label by $i=1,2$ respectively, corresponding to each leg of the contour $\Gamma$ attached to CFT$_i$ as in figure \ref{fig:contour}, 
\be
Z^{(i)} = c^{(i)}_A \widetilde{G}_{A}(\omega,z) + c^{(i)}_R\widetilde{G}_{R}(\omega,z). \label{ZiGreens}
\ee
The source (the non-normalisable data) associated to $Z^{(i)}$ is therefore, as read off at $z=0$,
\be
s^{(i)} = \lim_{z\to 0} z^{\Delta -d} Z^{(i)} =  c_A^{(i)} + c_R^{(i)}.\label{SKsources}
\ee
The two solutions \eqref{ZiGreens} are not independent, as we require them to be regular solutions on the contour $\Gamma$. They are therefore related by continuity along $\Gamma$ which goes around the branch points at $z=z_h$. This analytic continuation imposes relations between the four coefficients $c^{(i)}_{A,R}$,
\be
c^{(2)}_A = e^{\frac{\beta\omega}{2}}c^{(1)}_A,\qquad c^{(2)}_R = e^{-\frac{\beta\omega}{2}}c^{(1)}_R. \label{SKmatching}
\ee
Then \eqref{SKsources} and \eqref{SKmatching} are four relations which determine the four coefficients $c^{(i)}_{A,R}$ in terms of sources $s^{(i)}$ as follows,
\be \label{matchingconditions}
\begin{aligned}
c_A^{(1)} &= n\left(- s^{(1)} + e^\frac{\beta\omega}{2}s^{(2)}\right),\\
c_R^{(1)} &= n\left(e^{\beta\omega}s^{(1)}-e^\frac{\beta\omega}{2}s^{(2)}\right),\\
c_A^{(2)} &= n\left(- e^\frac{\beta\omega}{2}s^{(1)} + e^{\beta\omega}s^{(2)}\right),\\
c_R^{(2)} &= n\left(e^\frac{\beta\omega}{2}s^{(1)} -s^{(2)}\right),
\end{aligned}
\ee
where $n = (e^{\beta\omega} -1)^{-1}$ is the Bose-Einstein distribution function. These match the expressions (24a) and (24b) in \cite{Herzog:2002pc} when $\omega > 0$ with sources on the two-sided black hole, or, the prescription in \cite{Son:2009vu} with $\sigma = \beta/2$.

Finally we are ready to impose the normalisability condition. The fields take the following form near the boundary,
\bea
Z^{(1)} &=& s^{(1)}z^{d-\Delta} + \ldots \label{Znb1} \\
&& + \left(n(- s^{(1)} + e^\frac{\beta\omega}{2}s^{(2)}) G_A + n(e^{\beta\omega}s^{(1)}-e^\frac{\beta\omega}{2}s^{(2)})G_R\right)\frac{z^\Delta}{2\Delta -d} + \ldots \nonumber\\
Z^{(2)} &=& s^{(2)}z^{d-\Delta} + \ldots  \label{Znb2}\\
&& + \left(n(- e^\frac{\beta\omega}{2}s^{(1)} + e^{\beta\omega}s^{(2)}) G_A  + n(e^\frac{\beta\omega}{2}s^{(1)} -s^{(2)})G_R\right)\frac{z^\Delta}{2\Delta -d} + \ldots \nonumber
\eea
where $G_{A,R}$ are the advanced and retarded Green's function of the dual QFT, and ellipses denote higher order terms in the expansion that are determined by the data shown. Here we work at generic $\Delta > d/2$ and factors of $2\Delta - d$ come from holographic renormalisation \cite{Skenderis:2002wp}. Indeed, varying the one-point functions $v^{(j)}$ (i.e. $(-1)^{j+1}(2\Delta -d)$ times the coefficients of $z^{\Delta}$ in the expansions above) with respect to $s^{(j)}$ confirms the $\sigma = \beta/2$ Schwinger-Keldysh correlators,
\bea
G_{11} = \frac{\delta v^{(1)}}{\delta s^{(1)}}  &=& n(e^{\beta\omega}G_R-G_A),\\% = \re{G_R} + i \coth(\beta\omega/2) \im{G_R}\\
G_{12} = G_{21} = -\frac{\delta v^{(2)}}{\delta s^{(1)}} = -\frac{\delta v^{(1)}}{\delta s^{(2)}} &=& n e^{\frac{\beta\omega}{2}}(G_R-G_A),\label{G12}\\% = 2i \frac{e^{\frac{\beta\omega}{2}}}{e^{\beta\omega}-1} \im{G_R}\\
G_{22} = \frac{\delta v^{(2)}}{\delta s^{(2)}}  &=& n(G_R - e^{\beta\omega} G_A),% = -\re{G_R} + i \coth(\beta\omega/2) \im{G_R}
\eea
as in \cite{Herzog:2002pc}. Normalisability here is the condition that no sources are present, $s^{(i)} = 0$. First, let us set one of the boundary conditions, $s^{(1)} = 0$, and also normalise the subleading data there to unity. Then we have,
\bea
Z^{(1)} &=&  z^\Delta + \ldots,\\
Z^{(2)} &=& -\frac{(2\Delta-d)}{G_{12}}z^{d-\Delta} + \ldots  +  \frac{G_{22}}{G_{12}}z^\Delta + \ldots.
\eea
Therefore $Z^{(2)}$ is also normalisable only at the poles of $G_{12}$, which occur only at QNM and anti-QNM frequencies \cite{festuccia_thesis, Festuccia:2005pi, Dodelson:2023vrw}.\footnote{We included in this definition of QNMs and anti-QNMs some  special cases corresponding to so-called `pole-skipping' points, where QNMs at certain complex values of the quantum numbers $J$ have Matsubara frequencies, and appear as poles in $G_{12}$ despite not being poles of $G_{R/A}$ \cite{Dodelson:2023vrw}. In the $\Delta=2$ BTZ case, for instance, $\omega^+_{n=0,m=+i}=\omega^-_{n=0,m=-i}=-i$ in \eqref{wBTZ}, corresponding to the first Matsubara frequency, and the would-be pole in $G_R$ is indeed cancelled by a would-be zero. Similarly for \eqref{awBTZ} and $G_A$.} Hence QNMs and anti-QNMs are the only regular, normalisable eigenfunctions of $\mathcal{H}$ on the complex radial contour $\Gamma$. 

\section{Explicit evaluations on QNMs}\label{sec:examples}
The main result of this paper, the orthogonality of QNMs under \eqref{orthnorm}, was established for a general class of black holes in sections \ref{sec:orthogonality} and \ref{sec:efuncs}. However, it is still instructive to see explicit evaluations of $\braket{\Phi_a,\Phi_b}$ with $\Phi_{a,b}$ any two selected from the set of QNMs and anti-QNMs. Each choice takes the form,
\be
\Phi_{a,b}(t,z,\sigma) = e^{-i \omega_{a,b} t} Z_{a,b}(z) Y_{J_{a,b}}(\sigma),
\ee
where $Y_J$ are eigenfunctions of $\Delta_\sigma$ with quantum numbers $J_a,J_b$, enjoying an orthogonality relation,
\be
\int d\sigma_{d-1}Y^*_{J_a}(\sigma)Y_{J_b}(\sigma) = N_\sigma \delta_{J_aJ_b},
\ee
where in non-compact cases (on the plane) the $\delta$ is a Dirac delta.
Then, a useful intermediate result is,
\bea
\braket{\Phi_a,\Phi_b} &=& N_\sigma\,\delta_{J_aJ_b}\; e^{-i (\omega_b-\omega_a) t}(\omega_b + \omega_a)\, I_{ab}\label{productI}\\
I_{ab} &\equiv & \int_\Gamma dz\,\frac{Z_a(z)Z_b(z)}{z^{d-1}\sqrt{f(z)}\sqrt{g(z)}}.
\label{Idef}
\eea
Our focus is on evaluating the radial contour integral \eqref{Idef} for $J_a = J_b$.

The first step is to obtain the QNM and anti-QNM eigenfrequencies $\omega_a$. We do this by solving \eqref{Zeq} in $z\in[0,1]$, imposing ingoing behaviour at $z=1$ and normalisability at $z=0$ for QNMs. The radial operator are discretised using Chebyshev spectral methods, ensuring convergence with numerical resolution and digits of precision. Further details of this technique can be found, for example, in appendix C of \cite{Carballo:2024kbk}.

With the eigenfrequencies obtained, we construct the eigenfunctions on $\Gamma$, and $I_{ab}$. Recall that the eigenfunctions have branch point singularities at $z=z_h$, \eqref{nearhorizon}. This leads to a practical issue whereby going too close to $z=z_h$ leads to large numbers and reduces our ability to accurately evaluate $I_{ab}$ with numerics. To resolve this we define a new contour $\widetilde{\Gamma}$ which is simply the deformation of $\Gamma$, such that the endpoints are the same and no new singularities are crossed. We choose $\widetilde{\Gamma}$ to give $z=z_h$ a wide berth. By Cauchy's theorem the integral is the same, but the numerical approximations are improved.

$\widetilde{\Gamma}$ is taken to be piecewise, parametrised with a set of curves $z_i(\lambda)$ and $\lambda \in [0,1]$ we have
\be
I_{ab}  = \sum_i\int_0^1 \frac{Z_a(z_i(\lambda))Z_b(z_i(\lambda))}{z_i(\lambda)^{d-1}\sqrt{f(z_i(\lambda))}\sqrt{g(z_i(\lambda))}}\, z_i'(\lambda) d\lambda.
\ee
Then, given a pair of frequencies $\omega_a, \omega_b$ we wish to obtain $I_{ab}$ by solving the following set of three ODEs in $\lambda$, on each curve $z_i$,
\bea
F_i'(\lambda) &=& \frac{Z_a(z_i(\lambda))Z_b(z_i(\lambda))}{z_i(\lambda)^{d-1}\sqrt{f(z_i(\lambda))}\sqrt{g(z_i(\lambda))}}z_i'(\lambda), \label{numericsSpecification1}\\
\left(\mathcal{L}Z_a\right)(z_i(\lambda))&=& -i\omega_a^2Z_a(z_i(\lambda)),\label{numericsSpecification2}\\
\left(\mathcal{L}Z_b\right)(z_i(\lambda)) &=& -i\omega_b^2Z_b(z_i(\lambda)),\label{numericsSpecification3}
\eea
where we impose conditions $F_i(0) = 0$ on each curve. To start the process at $z=0 - i\epsilon$ i.e. at $\lambda = 0$ on the first curve, we impose $Z_a,Z_b \sim z^{\Delta}$ with arbitrary normalisation, which provides initial data for the first integration, $Z_a,Z_b,\partial_\lambda Z_a, \partial_\lambda Z_b$ at $\lambda = 0$. One then integrates \eqref{numericsSpecification1},\eqref{numericsSpecification2},\eqref{numericsSpecification3} along the first curve. At the end of the first curve, the values of $Z_a,Z_b,\partial_\lambda Z_a, \partial_\lambda Z_b$ are known there, and this gives the initial data for integrating along the next curve (with appropriate analytic continuation for the derivatives at corners). Thus continues until $\widetilde{\Gamma}$ is completed and then $I_{ab} = \sum_i F_i(1)$. To assess the numerical results it is useful to define a normalised version of the product \eqref{productI},
\be
(\Phi_a, \Phi_b) = \frac{|\braket{\Phi_a,\Phi_b}|}{||\Phi_a||\,||\Phi_b||}, \label{normprod}
\ee
where $||a|| \equiv \sqrt{|\braket{a,a}|}$.

As an example, we take Schwarzschild-AdS$_4$ black brane, with $K=0$, and we take $z_h=1$ so that $f(z) = g(z) = 1- z^3$. In this case we parameterise $\Gamma$ with the following three straight line segments, 
\be
z_i(\lambda) = z_i^{(a)} + \lambda \left(z_i^{(b)} - z_i^{(a)}\right),
\ee
forming a triangular contour that starts and ends at $z=\mp i\epsilon$ and avoids $z=1$,
\begin{align}
 z_1^{(a)} &= -i \epsilon, &  z_1^{(b)} &= \frac{3}{2} - \frac{i}{2},\\
 z_2^{(a)} &= \frac{3}{2} - \frac{i}{2},  & z_2^{(b)} &= \frac{3}{2} + \frac{i}{2},\\
 z_3^{(a)} &= \frac{3}{2} + \frac{i}{2}, & z_3^{(b)} &= i \epsilon.
\end{align}
The QNMs are arranged in a `christmas tree' structure with a $\omega \to -\omega^*$ symmetry. We label the QNMs in this tree as $n_\pm$, where $n$ labels the distance of the mode from the origin and $+$ denotes the $\re{\omega}>0$ branch, while $-$ denotes the $\re{\omega} <0$ branch, and similarly for anti-QNMs with a tilde, $\widetilde{n}_\pm$. The results from the numerical computation are displayed in table \ref{table:orthnumerics}.
Because of the prefactor $\omega_a + \omega_b$ in \eqref{productI} and the $\omega \to -\omega^*$ symmetry, the $n_\pm$ are automatically orthogonal to $\widetilde{n}_\mp$ giving some exact zeros in the table. 

\begin{table}[h!]
\small
\centering
\begin{tabular}{L||L|L|L|L|L|L|L|L}
 & 1_- & 1_+ & 2_- & 2_+ & \widetilde{1}_- & \widetilde{1}_+ & \widetilde{2}_- & \widetilde{2}_+\\
\hline\hline
1_- & 1 & 1\times 10^{-18} & 3\times 10^{-18} & 4\times 10^{-19} & 1\times 10^{-18} & 0 & 3\times 10^{-19} & 1\times 10^{-18}\\
1_+ & - & 1 & 2\times 10^{-17} & 6\times 10^{-19} & 0 & 1\times 10^{-18} & 2\times 10^{-19} & 2\times 10^{-17}\\
2_- & - & - & 1 & 1\times 10^{-16} & 2\times 10^{-17} & 1\times 10^{-18} & 8\times 10^{-17} & 0\\
2_+ & - & - & - & 1 & 2\times 10^{-19} & 3\times 10^{-19} & 0 & 8\times 10^{-17}\\
\widetilde{1}_- & - & - & - & - & 1 & 1\times 10^{-18} & 6\times 10^{-19} & 2\times 10^{-17}\\
\widetilde{1}_+ & - & - & - & - & - & 1 & 4\times 10^{-19} & 3\times 10^{-18}\\
\widetilde{2}_- & - & - & - & - & - & - & 1 & 1\times 10^{-16}\\
\widetilde{2}_+ & - & - & - & - & - & - & - & 1\\
\end{tabular}
\caption{The orthogonality relation \eqref{normprod} evaluated among all pairings between the first 4 QNMs and first 4 anti-QNMs of the Schwarzschild-AdS$_4$ black brane with $\Delta = 2$. We use numerical precision $40$ and $\epsilon = 10^{-7}$. We omit the lower triangular values as the product is manifestly symmetric. The off-diagonal entries become smaller as $\epsilon$ is reduced. Here $k=0$ is shown; similar level of confirmation was seen for values of $k \neq 0$.}
\label{table:orthnumerics}
\end{table}

\section{Discussion}\label{sec:discussion}
In this work we constructed an orthogonality relation between QNMs for a wide class of symmetric black holes in AdS$_{d+1}$. The key ingredients were a modification of the Klein-Gordon product with a $\mathcal{CPT}$ operator and a complex radial contour, $\Gamma$.

The $\mathcal{CPT}$ operator was required to map eigenfunctions of $\mathcal{H}$ with frequency $\omega$ into eigenfunctions with frequency $\omega^*$. This is because black holes present the peculiar scenario of having a formally self-adjoint Hamiltonian (on our contour $\Gamma$) with complex eigenvalues, due to the vanishing of Klein-Gordon norms. The $\mathcal{CPT}$ acts to diagonalise the product and produces the correct orthogonality relation. This result is not specific to black holes in AdS. Relatedly, complex eigenvalues with self-adjoint Hamiltonians also appear when rewriting perturbations of Schwarzschild as a supersymmetric quantum mechanics \cite{Bakas:2014kfa}. The requirement of pairing different left and right eigenfunctions was also seen in an AdS$_2$ example in \cite{Chen:2023hra} when deforming from normal modes in AdS$_2$ to the double-sided black hole. 

The choice of contour $\Gamma$ (see figure \ref{fig:contour}) is significant in three respects. Firstly, because it avoids the branch points at the horizon, the integral in the product is convergent as defined, rather than being a regulated version of something else. Secondly, there is no contribution from boundary terms at the horizon. This allows the Hamiltonian to be self-adjoint under this product; boundary terms at the horizon are responsible for the lack of self-adjointedness of the Hamiltonian in other products, preventing orthogonality \cite{Carballo:2024kbk}. Thirdly, $\Gamma$ connects two copies of the QFT on a thermal Schwinger-Keldysh contour as a natural construction in real-time AdS/CFT.

Let us comment on the relation of our work to previous orthogonality products in the literature \cite{Jafferis:2013qia, Witten:2001kn, Bousso:2001mw, Ng:2012xp, Crawley:2021ivb, Cotler:2023qwh, Green:2022htq, Cannizzaro:2023jle, PhysRevA.49.3057, Ching:1995rt, London:2023aeo, London:2023idh, Zhu:2023mzv}. In this paper, we provided a general explanation of the origin of QNM orthogonality in terms of left- and right-eigenfunctions of the Hamiltonian not addressed in previous works. In particular, our construction justifies from first principles the insertion of the discrete symmetry operators in one of the entries of a Klein-Gordon product, which appears also in some of the references listed above. Thus, we expect our framework to be applicable to any black hole or cosmological spacetime with QNMs, regardless of their asymptotics. Indeed, we have checked that our relation holds for scalar QNMs on the static patch of dS as well, and it is realised by taking the complex radial contour that encircles the cosmological horizon and connects the points $r=0\pm i\,\epsilon$, in agreement with the results in \cite{Jafferis:2013qia}.
The application of our relation to asymptotically flat spacetimes is more delicate due to the presence of null-infinity, and it has not been addressed directly in this work. However, we expect the same construction to remain valid with a suitable Schwinger-Keldysh contour connecting two copies of flat space asymptotically. This is potentially what is happening in \cite{Green:2022htq, London:2023aeo}, where a complexified contour was adopted for regulation purposes, but the connections with Schwinger-Keldysh and with the role of the $\mathcal{CPT}$ operator were not made.

An outstanding question is one of completeness -- whether QNMs and anti-QNMs form a suitable basis of regular, normalisable functions on the complex radial contour $\Gamma$. While the Hamiltonian is self-adjoint under \eqref{orthnorm}, this product lacks many basic properties to furnish us with a self-adjoint Sturm-Liouville problem. In particular the weight function on $\Gamma$ is complex, and product is not positive definite. We observe that QNMs alone do not form a basis, since the anti-QNMs are orthogonal to them under our product, however one may wish to specialise to functions that are regular on the future horizon, excluding anti-QNMs. We note there are counterexamples to completeness of QNMs only, on a single real copy of the exterior spacetime \cite{Warnick:2022hnc}. A related observation is that the two-sided holographic correlator, $G_{12}$, is constructed from only QNM frequencies and their complex conjugates \cite{Dodelson:2023vrw}.

Regardless of whether or not QNMs and anti-QNMs are complete, it is reasonable to expect that a sum of QNMs can approximate a solution at late times in the backgrounds we have considered \eqref{generalBG}. We expect that our product \eqref{orthnorm} will extract the relevant QNM expansion coefficients in this regime. Indeed, in asymptotically flat examples \cite{Green:2022htq}, projection using the appropriate orthogonality relation agrees with relevant `excitation coefficients' corresponding to the contribution of poles of the retarded Green's function \cite{Leaver, SunExcitation, Andersson:1996cm, BertiExcitation}. This physics can also be expressed in the language of a Keldysh spectral expansion \cite{Ansorg:2016ztf, Besson:2024adi} which can be constructed in a product-independent fashion \cite{Besson:2024adi}.

Concerning the generality of the product \eqref{orthnorm}. We based the construction of the product around a modification of the Klein-Gordon product. A similar construction can be made based around other products. For example, in appendix \ref{sec:energynorm} we present an analogous construction built around so-called `energy norms' that are second order functionals. The outcome is much the same. 

It is also natural to ask about the choice of bulk Hamiltonian, i.e. the choice of time slicing used in the bulk. We note that ingoing coordinates such as null or hyperboloidal slices \cite{Schmidt:1993rcx, Zenginoglu:2007jw, Dyatlov:2010hq, Bizon:2010mp,  Warnick:2013hba, PanossoMacedo:2018hab, Gajic:2019qdd, Bizon:2020qnd, PanossoMacedo:2023qzp} are set up to avoid QNM singularities on the past horizon. However, $\mathcal{CPT}$ exchanges QNMs with anti-QNMs, and so branch points will still appear inside any $\mathcal{CPT}$-modified product on ingoing slices. This would allow for a similar construction as presented here, but on an ingoing complex contour such as \cite{Glorioso:2018mmw}.

\section*{Acknowledgements}
It is a pleasure to thank Itamar Yaakov and Vaios Ziogas for discussions, and Alex Belin, Felix Haehl, Kostas Skenderis, and Vaios Ziogas for comments on a draft.
PA is supported by the Royal Society grant URF\textbackslash R\textbackslash 231002, `Dynamics of holographic field theories'.
JC is supported by the Royal Society grant RF\textbackslash ERE\textbackslash 210267.
BW is supported by a Royal Society University Research Fellowship and in part by the STFC consolidated grant `New Frontiers In Particle Physics, Cosmology And Gravity'. 
\appendix

\section{Further details of the BTZ example} \label{sec:BTZ}
The QNMs for a complex scalar obeying $\Box \Phi = \Delta(\Delta-2)\Phi$ on the BTZ background \eqref{BTZmetric} are given as follows,
\bea
\Phi_{nm}^\pm &=&  Z_{nm}^\pm(z)\, e^{im\varphi}\,e^{-i \omega^\pm_{nm}t}, \label{QNMBTZ}\\
Z_{nm}^\pm 
&=& z^\Delta\left(z^2-1\right)^{\mp\frac{i m}{2}-\frac{\Delta }{2}} \, {}_2F_1\left(-n,\pm i m+n+\Delta ;\Delta ;\frac{z^2}{z^2-1}\right),\label{hypBTZ}\\
\omega^\pm_{nm} &=& \pm m-i(2n + \Delta),\label{wBTZ}
\eea
while the anti-QNMs are given by
\bea
\widetilde{\Phi}_{nm}^\pm &=&  \widetilde{Z}_{nm}^\pm(z)\, e^{im\varphi}\,e^{-i \widetilde{\omega}^\pm_{nm}t},\label{aQNMBTZ}\\
\widetilde{Z}_{nm}^\pm(z) &=& z^\Delta\left(z^2-1\right)^{\pm\frac{i m}{2}-\frac{\Delta }{2}} \, {}_2F_1\left(-n,\mp i m+n+\Delta ;\Delta ;\frac{z^2}{z^2-1}\right)\label{hypaBTZ}\\
&=& Z_{n(-m)}^\pm(z) = Z_{nm}^\mp(z)\\
\widetilde{\omega}^\pm_{nm} &=& \pm m+i(2n + \Delta).\label{awBTZ}
\eea
Note that $\mathcal{CPT}\widetilde{\Phi}_{nm}^\pm = \Phi_{nm}^\pm$. In this case, the radial part of the product defined in \eqref{Idef}, $I_{ab}$, when adorned with all relevant labels, becomes
\begin{equation}
I_{n,n',m}^{\pm,\pm'}\equiv \int_\Gamma dz \frac{1}{z(1-z^2)}\,Z_{nm}^\pm(z)\, Z_{n'm}^{\pm'}(z),
\end{equation}
and we therefore have 
\bea
\braket{\Phi_{nm}^\pm,\Phi_{n'm'}^{\pm'}} &=& 2\pi\delta_{mm'}\,(\omega^{\pm'}_{n'm} + \omega^{\pm}_{nm}) e^{-i (\omega^{\pm'}_{n'm}- \omega^\pm_{nm})t} \,I_{n,n',m}^{\pm,\pm'}, \label{integralBTZ}\\
\braket{\widetilde{\Phi}_{nm}^\pm,\widetilde{\Phi}_{n'm'}^{\pm'}} &=&2\pi\delta_{mm'}\,(\widetilde{\omega}^{\pm'}_{n'm} + \widetilde{\omega}^{\pm}_{nm}) e^{-i (\widetilde{\omega}^{\pm'}_{n'm}- \widetilde{\omega}^\pm_{nm})t}\,I_{n,n',m}^{\mp\mp'},\\
\braket{\Phi_{nm}^\pm,\widetilde{\Phi}_{n'm'}^{\pm'}} &=&2\pi\delta_{mm'}\,(\widetilde{\omega}^{\pm'}_{n'm} + \omega^{\pm}_{nm}) e^{-i (\widetilde{\omega}^{\pm'}_{n'm}- \omega^\pm_{nm})t}\,I_{n,n',m}^{\pm,\mp'}. \label{PrelimBTZOverlap}
\eea
Thus all products are known once $I_{n,n',m}^{\pm,\pm'}$ is evaluated.
To evaluate $I_{n,n',m}^{\pm,\pm'}$ we note that the hypergeometeric functions in \eqref{hypBTZ} can be written as finite sums,
\begin{equation}
{}_2F_1\left(-n,\pm i m+n+\Delta ;\Delta ;\frac{z^2}{z^2-1}\right)=\sum_{s=0}^n(-1)^s\frac{n!}{s!\,(n-s)!}\frac{(\pm i m+n+\Delta)_s}{(\Delta)_s}\left(\frac{z^2}{z^2-1}\right)^s,
\end{equation}
Then, moving these sums outside the integral we obtain,
\bea
I_{n,n',m}^{\pm\pm'}=\sum_{s=0}^n\sum_{s'=0}^{n'}(-1)^{s+s'}\frac{n!\,(\pm i m+n+\Delta)_s}{s!\,(n-s)!\,(\Delta)_s}\frac{n'!\,(\pm' i m+n'+\Delta)_{s'}}{s'!\,(n'-s')!\,(\Delta)_{s'}}\,\widetilde{I}_{s,s',m}^{\pm\pm'}, \label{BTZdoublesum}
\eea
where we have defined
\begin{equation}
\begin{aligned}
\widetilde{I}_{s,s',m}^{\pm\pm'}&\equiv\int_\Gamma z^{2\Delta+2 s+2 s'-1} \left(z^2-1\right)^{-\Delta\mp \frac{i m}{2}\mp'\frac{i m}{2}-s-s'-1}\,dz\\
&=\frac{i \pi  \Gamma \left(s+s'+\Delta\right)}{\Gamma \left(\mp\frac{i m}{2} \mp'\frac{i m}{2}\right) \Gamma \left(1\pm\frac{i m}{2}\pm'\frac{i m}{2}+\Delta+s+s'\right)}.
\end{aligned}
\end{equation}
We note the integral above is a special case of the following useful result,
\be
\int_\Gamma z^a \left(z^2-1\right)^b\,dz = \frac{i\pi \Gamma\left(\frac{1+a}{2}\right)}{\Gamma(-b)\Gamma\left(b + \frac{3+a}{2}\right)},\qquad (\re{a}>-1).
\ee
Performing the summations in \eqref{BTZdoublesum}, we find
\be
I_{n,n',m}^{\pm\pm'}=\frac{i \pi  (-1)^{n'-1}\, n! \,\Gamma \left(\Delta\right)\left(\Delta \pm \frac{i m}{2}\pm' \frac{i m}{2}+n+n'\right)^{-1}}{\Gamma (n-n'+1) \Gamma (-n+n'+1)  \Gamma \left(\mp \frac{i m}{2}\mp' \frac{i m}{2}-n'\right)\,\left(\Delta\right)_{n'}\, \Gamma \left(n\pm \frac{i m}{2}\pm' \frac{i m}{2}+\Delta\right)}.
\ee
Because of the presence of the $\Gamma$-functions $\Gamma (n-n'+1), \Gamma (-n+n'+1)$ in the denominator, the result vanishes unless $n=n'$, thus,
\begin{equation}
I_{n,n',m}^{\pm\pm'}=\frac{i \pi  (-1)^{n-1}\, n! \,\Gamma \left(\Delta\right)\left(\Delta \pm \frac{i m}{2}\pm' \frac{i m}{2}+2n\right)^{-1}}{\Gamma \left(\mp \frac{i m}{2}\mp' \frac{i m}{2}-n\right)\,\left(\Delta\right)_{n}\, \Gamma \left(n\pm \frac{i m}{2}\pm' \frac{i m}{2}+\Delta\right)}\delta_{nn'}.
\end{equation}
Moreover, because of the presence of the $\Gamma$-function $\Gamma \left(\mp \frac{i m}{2}\mp' \frac{i m}{2}-n\right)$ in the denominator, we see that when $\pm\ne\pm'$ the result vanishes, as well as the case $m=0$. Hence,
\begin{equation}
\begin{aligned}
&I_{n,n',m}^{\pm\pm'}=\frac{i \pi  (-1)^{n-1}\, n! \,\Gamma \left(\Delta\right)}{\Gamma \left(\mp i m-n\right)\, \Gamma \left(n\pm i m+\Delta\right)\,\left(\Delta\right)_{n}\,\left(\Delta \pm i m+2n\right)}\delta_{nn'}\delta_{\pm\pm'}.
\end{aligned}
\end{equation}
Through \eqref{PrelimBTZOverlap}, this result implies that $\braket{\Phi_{nm}^\pm,\widetilde{\Phi}_{n'm'}^{\pm'}} = 0$; the Kronecker deltas in $I_{n,n',m}^{\pm,\mp'}$ enforce $n=n'$ and $\pm=\mp'$, whereupon the prefactor vanishes, $\widetilde{\omega}_{nm}^{\mp}+\omega_{nm}^{\pm}=0$. Thus we obtain the main BTZ results \eqref{BTZoverlap1}, \eqref{BTZoverlap2}, \eqref{BTZoverlap3}.

\section{$\mathcal{CPT}$ energy norms}\label{sec:energynorm}
In the main text we have presented a $\mathcal{CPT}$ modification of the Klein-Gordon product on a complex contour $\Gamma$, under which QNMs are orthogonal. QNM orthogonality under this new product ultimately stemmed from orthogonality between right and left eigenfunctions of $\mathcal{H}$ on $\Gamma$ in the original Klein-Gordon product \eqref{leftright}, using that $\mathcal{H}^{\dagger_{\text{KG}}}=\mathcal{H}$, and the mapping between these pairings that $\mathcal{CPT}$ implements. Thus, analogous QNM orthogonality relations can be constructed from a different starting product $\braket{\cdot,\cdot}_\Omega$ that also satisfies $\mathcal{H}^{\dagger_\Omega}=\mathcal{H}$ on $\Gamma$.  In this appendix, we present one such an example where the starting point is the so-called `energy norm', often used in the context of the black hole pseudospectrum \cite{Jaramillo:2020tuu, Gasperin:2021kfv}, see also \cite{Driscoll, SchmidRev, TrefethenHSWE}. We make the analogous modifications of it, and show it serves the same role.

Consider the energy of a complex scalar $\Phi$ on a fixed $t$ slice,
\be
E = \int d\Sigma^\mu T_{\mu t},
\ee
where
\be
\label{eq:energymomentum_Phi}
T_{\mu\nu} = \frac{1}{2}\nabla_{\mu} \Phi^*\nabla_{\nu}\Phi + \frac{1}{2}\nabla_{\nu}\Phi^*\nabla_{\mu}\Phi - g_{\mu\nu}\left(\frac{1}{2}\nabla_{\alpha}\Phi^*\nabla^{\alpha}\Phi + \frac{1}{2}m^2\Phi^*\Phi \right).
\ee
This is a quadratic functional that inspires the following energy norm, here written at fixed quantum number for the transverse eigenfunctions, with eigenvalue of $\Delta_\sigma Y(\sigma) = -\lambda Y(\sigma)$,
\be
\braket{a,b}_E = \frac{\text{vol}_\sigma}{2}\int_\mathcal{C} \frac{dz}{z^{d-1}\sqrt{f}\sqrt{g}} \left(\partial_t a^* \partial_t b + fg \partial_z a^* \partial_z b + \left(\frac{m^2}{z^2} + \lambda\right)fa^*b\right),
\ee
such that $E[a]= \braket{a,a}_E$. Then, we can define the modified product following \eqref{orthnorm},
\be
\braket{a,b}_{E,\mathcal{CPT}} = \braket{\mathcal{CPT}a,b}_E.
\ee
Similar to what happens for the product \eqref{orthnorm}, here one can also show that
\bea
\braket{a,(\mathcal{H}- \mathcal{H}^\dagger)b}_{E,\mathcal{CPT}} &=& \frac{i \text{vol}_\sigma}{2}\left[z^{1-d}\sqrt{f}\sqrt{g}\left(\partial_t (\mathcal{T}a)\partial_z b + \partial_z (\mathcal{T}a) \partial_t b\right)\right]_{z=0-i\epsilon}^{z=0+i\epsilon}\\
&=& 0,
\eea
from which orthogonality follows. We have also verified this result with numerical tests. This shows that orthogonality is not a special property of the modified Klein-Gordon product, and instead emphasises the importance of the $\mathcal{CPT}$ and complex contour modifications.

\bibliographystyle{ytphys}
\bibliography{refs}

\end{document}